%% file: Paper.tex
\documentclass[12pt]{article}

\usepackage{multibib}
\newcites{SI}{References for Supporting Information}
\usepackage[colorlinks=true,linkcolor=blue,citecolor=blue]{hyperref}
\usepackage{PRIMEarxiv}
\usepackage[utf8]{inputenc}
\usepackage[T1]{fontenc}
\usepackage{cmbright}
\usepackage{url}
\usepackage{booktabs}
\usepackage{amsfonts}
\usepackage{nicefrac}
\usepackage{microtype}
\usepackage{lipsum}
\usepackage{fancyhdr}
\usepackage{graphicx}
\usepackage{ragged2e}
\usepackage[table]{xcolor}
\usepackage{float}
\usepackage{caption}
\graphicspath{{media/}}
\usepackage{amsmath}
\usepackage{cite}
\usepackage{hyphenat}

\makeatletter
\renewcommand{\@maketitle}{
  \begin{center}
    {\LARGE \bfseries \@title \par}
    \vskip 1em
    {\large \@author}
    \vskip 1em
  \end{center}
}
\makeatother

\title{IS GELATION A SINGULARITY OR A FLOW-INDUCED INSTABILITY?}

\author{
\textbf{Manuel Dedola}$^{\dag}$,
\textbf{Ludovico Cademartiri}$^{\dag *}$ \\[1em]
$^{\dag}$Department of Chemistry, Life Sciences and Environmental Sustainability,\\
University of Parma, Parco Area delle Scienze 17 A, Parma, Italy.\\[0.5em]
$^{*}$Corresponding author: \href{mailto:ludovico.cademartiri@unipr.it}{ludovico.cademartiri@unipr.it}
}

\begin{document}
\maketitle

\RaggedRight 

\begin{abstract}
Gelation in the Smoluchowski coagulation equation is commonly interpreted as a finite-time singularity marked by mass loss or moment divergence. We instead characterize gelation as a loss of dynamical stability of the Smoluchowski flow, quantified through the time-dependent spectrum of the Jacobian along the evolving aggregation dynamics. Studying homogeneous kernels $K(i,j)=(ij)^{\alpha}$ together with the classical Smoluchowski, we show that gelation is consistently preceded by the appearance of positive real eigenvalues, indicating a loss of local dynamical stability. While non-gelling kernels exhibit only transient finite-size effects, gelling kernels display persistent spectral destabilization associated with macroscopic gel formation. Our results identify gelation as a genuine dynamical instability of the Smoluchowski flow.

\end{abstract}

\section{Introduction}

The Smoluchowski coagulation equation is a cornerstone model for irreversible aggregation processes, with applications ranging from colloidal aggregation and polymerization to cloud formation and network growth \cite{smoluchowski1916,friedlander2000}. It describes the time evolution of the cluster size distribution through binary coagulation events governed by an interaction kernel $K(i,j)$, encoding the rate at which clusters of sizes $i$ and $j$ merge.

One of the most striking phenomena arising in this framework is \textit{kinetic} gelation, namely the formation of a macroscopic cluster that absorbs a finite fraction of the total mass in finite time. We emphasize that this kinetics runaway growth - distinct from density-driven jamming or percolation \cite{aldous1999} - is an intrinsic property of the aggregation kernel in the dilute limit. Since Smoluchowski’s original work \cite{smoluchowski1916}, gelation has traditionally been characterized through moment-based criteria, such as the divergence of higher-order moments or the loss of mass conservation beyond a critical time $t_g$ \cite{leyvraz2003,aldous1999,lushnikov1973,ernst1984, hendriks1983}. For homogeneous kernels of the form $K(i,j)=(ij)^{\alpha}$, classical theory predicts gelation for $\alpha>1/2$, while subcritical kernels remain mass conserving for all times \cite{aldous1999,vanDongen1986,vanDongen1988}.\\
While these criteria provide a precise characterization of when classical solutions break down, they offer limited insight into the dynamical mechanism underlying gelation. In particular, moment divergence is a global diagnostic that does not directly address whether gelation corresponds to a loss of stability of the aggregation dynamics itself.

Despite this well-established theoretical picture, the dynamical nature of gelation remains conceptually subtle. In most treatments, gelation is described as a singular breakdown of the classical solution, accompanied by divergent moments or the introduction of a separate gel phase \cite{aldous1999}. While mathematically precise, this viewpoint provides limited insight into the \emph{dynamical mechanism} by which the aggregation process loses stability. Moreover, numerical studies have long reported finite-size effects, transient mass loss, and apparent gelation-like behavior even for kernels that are theoretically non-gelling.
This raises a fundamental question:\emph{Is gelation merely a singularity of moments, or does it correspond to a genuine dynamical instability of the Smoluchowski flow?}
In this work, we address this question from a dynamical systems perspective. Rather than focusing on global observables such as mass or moments, we analyze the \emph{local stability properties} of the full Smoluchowski equation, viewed as an infinite-dimensional nonlinear ODE system. Specifically, we compute the Jacobian of the coagulation dynamics \emph{along its evolving trajectory} and track the time-dependent spectrum of linear perturbations.

Although the Smoluchowski equation admits no nontrivial steady states prior to gelation, its time-dependent flow is well defined, and the instantaneous Jacobian spectrum provides direct information on the amplification or decay of perturbations. The emergence of eigenvalues with positive real parts signals a loss of local stability, independently of moment divergence or mass conservation.

We apply this framework to homogeneous multiplicative kernels in both subcritical and supercritical regimes, to the classical Smoluchowski kernel, and to a non-gelling additive kernel used as a control case. Our results show that gelation is consistently preceded by a persistent spectral destabilization of the Jacobian, while non-gelling kernels exhibit at most transient, finite-size instabilities that vanish in the large-cutoff limit.

These findings provide strong evidence that gelation should be interpreted not only as a singularity of moments, but as a \emph{dynamical instability of the aggregation flow itself}. By reframing gelation in terms of stability theory, this work establishes a direct conceptual link between classical coagulation theory and modern dynamical systems analysis, and introduces a robust diagnostic tool for identifying gelation beyond traditional moment-based criteria.

\section{Model and Stability Framework}

We consider the discrete Smoluchowski coagulation equation for the cluster size distribution $c_p(t)$ \cite{drake1972}:

\begin{equation}
\dot{c}_p=\frac{1}{2} \sum_{i+j=p} K_{i j} c_ic_j - c_p \sum_{j=1}^{\infty} K_{p j}c_j
\end{equation}

where $c_p(t)$ denotes the concentration of clusters of size $p$ at time $t$, and $K_{ij}$ is the coagulation kernel. Throughout this work, concentrations are measured in units of number density, and time is rescaled so that the overall collision rate is unity.

\subsection{Kernels Considered}
We focus on two representative classes of coagulation kernels to test the robustness of the spectral stability analysis.

\paragraph{Multiplicative homogeneous kernels.}
The primary family under investigation is the multiplicative kernel:

\begin{equation}
K_{ij} = (ij)^{\alpha}
\end{equation}

This class exhibits a sharp transition in the kinetic behavior: classical theory predicts gelation (mass loss in finite time) for $\alpha > 1/2$, while mass is conserved for $\alpha \le 1/2$ \cite{hendriks1983,leyvraz2003}. This allows us to sweep across the critical threshold and monitor the Jacobian spectrum.

\paragraph{Smoluchowski diffusion-limited kernel.}
As a physically motivated non-gelling control, we consider the classical kernel for Brownian coagulation of fractal aggregates in the continuum regime \cite{sandkuhler2005}:

\begin{equation}
K_{ij} =\frac{1}{4}(i^{-1/D_f}+j^{-1/D_f})(i^{1/D_f}+j^{1/D_f})
\end{equation}

where $D_f$ represents the fractal dimension of the aggregates. In our simulations, we fix $D_f = 1.8$, a typical value for diffusion-limited cluster aggregation (DLCA) \cite{lazzari2016}. The Smoluchowski kernel is homogeneous ($K(\omega i, \omega j)=\omega K(i,j)$) of degree $\omega = 0$. According to the rigorous classification by van Dongen and Ernst \cite{aldous1999,vanDongen1986}, homogeneous kernels with $\omega \le 1$ obey mass conservation for all times and do not undergo kinetic gelation. 

\subsection{Finite-dimensional truncation}

To make the problem numerically tractable, we introduce a cutoff $p$ and restrict cluster sizes to $1 \le i \le p$. This yields a finite-dimensional nonlinear ODE system,

\begin{equation}
\dot{\mathbf{c}} = \mathbf{F}(\mathbf{c}),
\qquad
\mathbf{c}(t) \in \mathbb{R}^p,
\end{equation}

which converges to the infinite-dimensional dynamics as $p \to \infty$ prior to gelation. All numerical results are obtained by systematically increasing $p$ to assess finite-size effects.

\subsection{Linearization along the trajectory}

Unlike classical stability analyses based on fixed points, the Smoluchowski equation admits no nontrivial steady states in the pre-gelation regime. We therefore adopt a \emph{time-dependent linearization} approach. Given a solution $\mathbf{c}(t)$ of Eq. \eqref{eq:finite_ode}, we define the Jacobian matrix \cite{goldhirsch1987stability,Cod}:

\begin{equation}
\textbf{J}(t)=\left. \frac{\partial \mathbf{F}}{\partial \mathbf{c}} \right|_{\mathbf{c}(t)},
\end{equation}

which governs the evolution of infinitesimal perturbations $\delta \mathbf{c}$ via:

\begin{equation}
\frac{d}{dt} \delta \mathbf{c} = \textbf{J}(t)\, \delta \mathbf{c}.
\end{equation}

At each time $t$, the spectrum of $\textbf{J}(t)$ provides instantaneous information on the local stability of the flow. In particular, the presence of eigenvalues with positive real part indicates exponential growth of perturbations and a loss of local stability of the aggregation dynamics.

\subsection{Explicit Jacobian Form}
The linearization of the discrete Smoluchowski equation yields the exact analytical form of the Jacobian operator elements $J_{ij}$:

\begin{equation}
J_{ij}(t)=\frac{\partial \dot{c}_i}{\partial c_j} = \underbrace{K{j, i-j} c_{i-j}(t) \cdot \mathbb{I}_{(j < i)}}_{\text{Generation (Positive Feedback)}} - \underbrace{K_{ij} c_i(t)}_{\text{Dissipation (Negative Feedback)}} - \underbrace{\delta{ij} \sum_{k=1}^{p} K_{ik} c_k(t)}_{\text{Dissipation (Negative Feedback)}} 
\end{equation}

where $\delta_{ij}$ is the Kronecker delta and $\mathbb{I}_{(\cdot)}$ is the indicator function, which equals 1 if the condition is met and 0 otherwise. We emphasize that the Jacobian spectrum is evaluated along the trajectory $\textbf{c}(t)$ and therefore does not define asymptotic stability in the strict dynamical systems sense. However, the persistence of positive real eigenvalues over extended time intervals provides a robust indicator of a genuine dynamical destabilization of the flow.

\subsection{Spectral indicator of gelation}

We define the maximal instantaneous growth rate: 
 
\begin{equation}
\lambda_{\max}(t) = \max \left\{ \mathrm{Re}(\lambda) : \lambda \in \sigma(\textbf{J}(t)) \right\},
\end{equation}

In practice, we interpret $\lambda_{max}(t)$ as an instantaneous growth rate that characterizes the local tendency of the aggregation dynamics to amplify or damp perturbations and use its long-time behavior as a diagnostic for gelation. A persistent positive $\lambda_{\max}(t)$ signals sustained destabilization of the flow, while transient positive excursions that vanish as $p$ increases are interpreted as finite-size effects.

Importantly, this spectral criterion does not rely on moment divergence or mass conservation arguments. Instead, it probes the intrinsic dynamical stability of the PBE flow itself, allowing us to distinguish genuine gelation from numerical or finite-size artifacts.

In the following section, we apply this framework to gelling and non-gelling kernels and demonstrate that spectral destabilization provides a sharp and robust signature of gelation.

\begin{figure}[H]
    \centering
\includegraphics[width=0.7\linewidth]{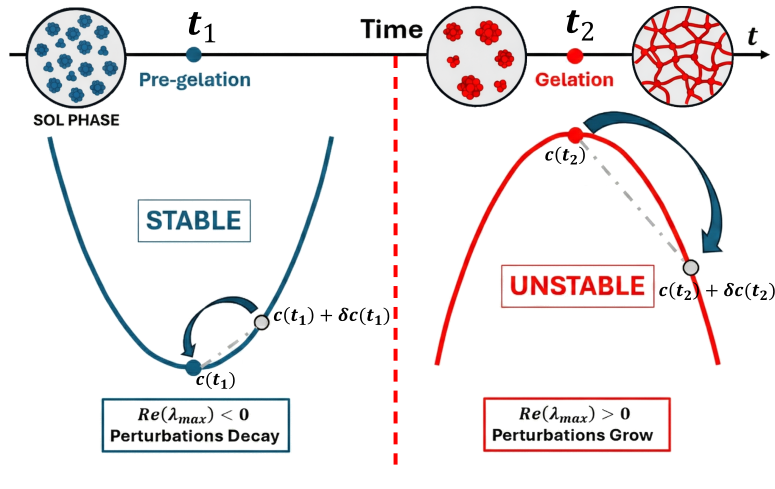}
    \caption{\textbf{Physical and dynamical signature of gelation}.
\textbf{Left}: Structural transition from dispersed clusters (Sol) to a macroscopic network (Gel). \textbf{Right}: Corresponding stability landscape. The sol-state is linearly stable ($\text{Re}(\lambda_{\max}) < 0$), whereas the gelation point marks the onset of a dynamical instability ($\text{Re}(\lambda_{\max}) > 0$), causing perturbations $\delta c$ to grow exponentially.}
    \label{Fig_1}
\end{figure}

\section{Computational Design}
The numerical framework was developed to solve the truncated Smoluchowski system and track its spectral evolution in real-time.
\subsection{Numerical Integration and Truncation}
We implement the discrete Smoluchowski equation by introducing a finite-size cutoff $p$, restricting cluster sizes to $1 \le i \le p$. The resulting system of $p$ nonlinear ODEs is integrated using the MATLAB$^{\circledR{}}$ ODE solver (ode15s).\\
The system is initialized as a monodisperse suspension $c_1(0)=1.0$ (normalized concentration, $c_p=N_p/N_0$). For each integration step, the coagulation rates $K_{ij}$ are pre-calculated as a $p \times p$ matrix to optimize the evaluation of the gain and loss terms. 

\subsubsection{Comparative Scaling and Convergence}
To differentiate between physical gelation and numerical artifacts, the simulation protocol includes:
\subsubsection{Kernel Comparison}
Systematic testing of gelling multiplicative kernels (${\alpha > 1/2}$) against non-gelling additive and diffusion-limited controls.

\subsubsection{Cutoff Scaling}

To distinguish genuine dynamical features from finite-size artifacts, we performed a scaling analysis by varying the system size cutoff $p$ (e.g., from $500$ to $2000$). This ensures that the observed spectral destabilization is a persistent feature of the infinite-dimensional flow rather than a transient effect of the truncation. Based on the convergence analysis (see Supporting Information, Fig. S1), we set $p=1000$ for the main results, as this value was found sufficient to capture the asymptotic behavior of the sol-gel transition while maintaining computational efficiency.

\subsubsection{Spectral Implementation}
Unlike traditional post-processing, our algorithm computes the Jacobian matrix $\mathbf{J}(t)$ at each discrete time step $t_n$ along the trajectory. The matrix $\mathbf{J}$ is constructed by explicitly evaluating the partial derivatives $\partial \dot{c}_i / \partial c_j$ for the current concentration vector $\mathbf{c}(t)$.
We perform a full eigendecomposition to extract the spectrum $\sigma(\mathbf{J}(t))$. This allows for the simultaneous tracking of the dominant growth rate $Re(\lambda_{max})$ and the overall distribution of eigenvalues in the complex plane.

\subsubsection{Observables and Statistical Indicators}
To complement the spectral analysis, the framework tracks the evolution of global observables that characterize the aggregation state and the distribution's complexity:

\begin{enumerate}
    \item \textbf{Moments Analysis}: We compute the first moment ($M_1 = \sum_{i=1}^p i c_i$) to monitor total mass conservation and identify the critical time $t_g$ when mass loss occurs in supercritical regimes. Simultaneously, the second moment ($M_2 = \sum_{i=1}^p i^2 c_i$) is tracked as the standard indicator of polydispersity, signaling the traditional gelation singularity through its divergence.
    \item \textbf{Shannon Entropy ($H$)}: To quantify the informational evolution of the cluster size distribution, we calculate the instantaneous Shannon entropy:$$H(t) = -\sum_{i=1}^p P_i \ln P_i$$where $P_i = c_i / \sum c_i$ represents the normalized probability of finding a cluster of size $i$. This metric provides a measure of the "disorder" of the distribution as it shifts from a monodisperse monomer state to a polydisperse sol and, eventually, to the gel phase.
\end{enumerate}

We first examine the dynamical stability of the aggregation process by comparing two representative multiplicative kernels, $K_{ij}=(ij)^\alpha$, in the subcritical ($\alpha=0.2$) and supercritical ($\alpha=0.8$) regimes.\\

\begin{figure}[H]
    \centering
    \includegraphics[width=0.8\linewidth]{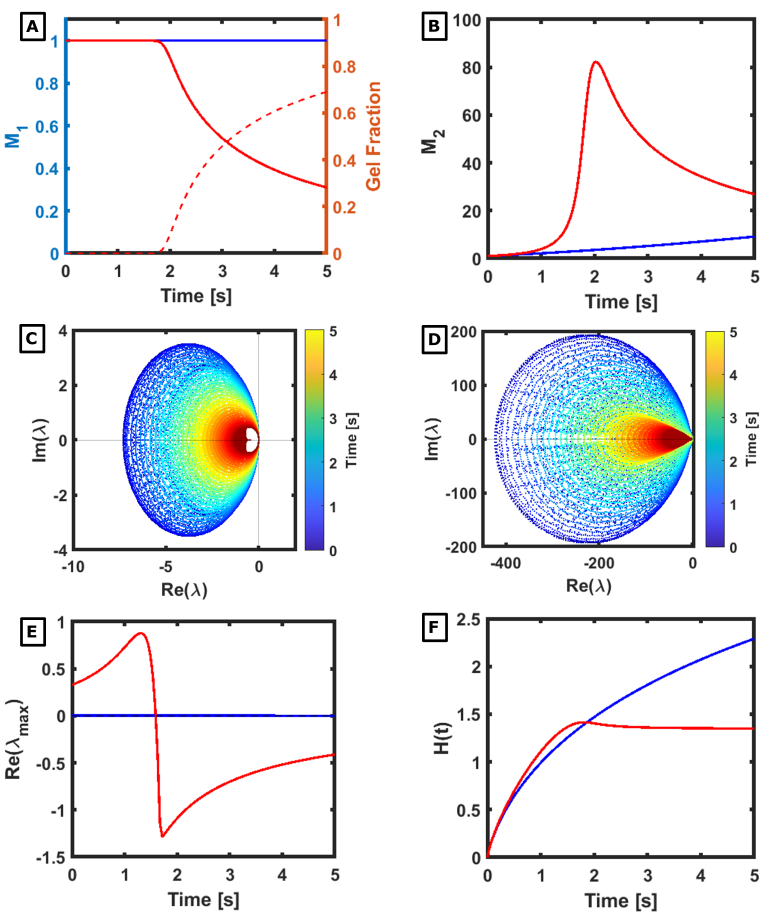}
    \caption{\textbf{Dynamics of the non-gelling and jelling.} (A) Evolution of the first moment ($M_1$) for $\alpha = 0.2$ (blue line) and $\alpha = 0.8$ (red line), and the gel fraction at $\alpha = 0.8$ (red dashed line). (B) Second moment ($M_2$) evolution, indicating polydispersity. Eigenvalue maps in the complex plane ($Im(\lambda)$ vs $Re(\lambda)$) over time (color-coded) for $\alpha=0.2$ (C) and $\alpha=0.8$ (D).
    (E) Real part of the maximum eigenvalue ($Re(\lambda_{max})$). The transition from positive to negative values indicates a shift in the dynamic stability of the population during the phase transition. (F) Evolution of structural parameter $H(t)$.}
    \label{Fig_2}
\end{figure}

In the non-gelling regime ($\alpha=0.2$), the system remains in the sol phase indefinitely. As shown in Figure \ref{Fig_2} (panels corresponding to $\alpha=0.2$), the first moment $M_1$ is strictly conserved, and the second moment $M_2$ grows monotonically without divergence. The maximum real eigenvalue of the Jacobian remains vanishingly small ($Re(\lambda_{max}) \approx 0$) throughout the evolution. This behavior reflects the neutral stability inherent to the Smoluchowski flow under mass-conserving kernels and indicates that perturbations introduced into the sol phase decay or propagate without amplification.\\

The dynamics change drastically in the gelling regime ($\alpha=0.8$). As the system approaches the critical gelation time $t_g$, the second moment $M_2$ diverges, and the first moment $M_1$ subsequently exhibits mass loss (Figure \ref{Fig_2}A-B). Coinciding with this singularity (Figure \ref{Fig_2}E, red line), $Re(\lambda_{max})$ - which is already positive - undergoes a macroscopic excursion to a maximum peak, signaling the height of the dynamical instability. Following the formation of the gel, $Re(\lambda_{max})$ drops sharply into negative values. This identifies the gel phase as a dynamically distinct state where the remaining sol population returns to a locally stable regime.\\
The structural complexity is tracked via Shannon entropy $H(t)$ (Figure \ref{Fig_2}F), Subcritical condition($\alpha=0.2$, blue): $H(t)$ grows monotonically as the distribution broadens indefinitely toward a polydisperse sol. Supercritical condition ($\alpha=0.8$, red): $H(t)$ saturates and decreases post-gelation as mass "orders" into a macroscopic cluster, limiting the informational disorder of the remaining.

\section{Results and Discussion}

\begin{figure}[H]
    \centering
    \includegraphics[width=0.9\linewidth]{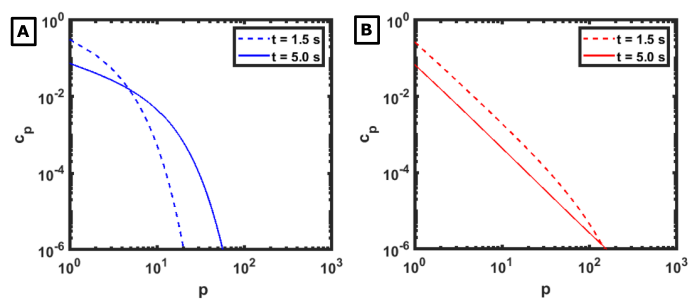}
    \caption{Cluster size distributions $c_p$ at intermediate ($t=1.5$) and late ($t=5.0$) times, comparing the non-gelling ($\alpha=0.2$) and gelling ($\alpha=0.8$) regimes.}
    \label{Fig_4}
\end{figure}

The structural difference between the two regimes is further highlighted by the cluster size distributions $c_p(t)$ (Figure \ref{Fig_4}). For $\alpha=0.2$, the distribution broadens over time but retains an exponential cutoff, characteristic of non-gelling systems. In contrast, for $\alpha=0.8$, the distribution develops a power-law tail that extends to the cutoff scale $p$ as $t \to t_g$, facilitating the formation of the macroscopic cluster.

To ensure that the observed spectral instability is a specific signature of gelation and not a numerical artifact of growing cluster sizes, we analyze the physically motivated Smoluchowski diffusion-limited kernel ($D_f=1.8$). This kernel describes realistic aggregation in the continuum limit and is rigorously known to be mass-conserving ($\omega=1$).\\

The results, summarized in Figure \ref{Fig_4}, provide an important negative control. Despite the significant evolution of the system - evidenced by the growth of polydispersity ($M_2$) and Shannon entropy ($H$) - the spectral stability is perfectly preserved. The Jacobian spectrum remains strictly confined to the left half-plane (Figure \ref{Fig_4}D), and $\text{Re}(\lambda_{\text{max}})$ remains vanishingly small ($\approx 0$) for all times (Figure \ref{Fig_4}C). Unlike the gelling case, there is no "leakage" of eigenvalues into the unstable right half-plane. 

This confirms that our spectral framework is robust: it correctly detects instability in gelling kernels while avoiding false positives in physically growing but non-gelling systems. The positive eigenvalues observed for $\alpha > 1/2$ are therefore not trivial consequences of cluster growth, but markers of the catastrophic breakdown of the mean-field description associated with gelation.

\begin{figure}[H]
    \centering
    \includegraphics[width=0.80\linewidth]{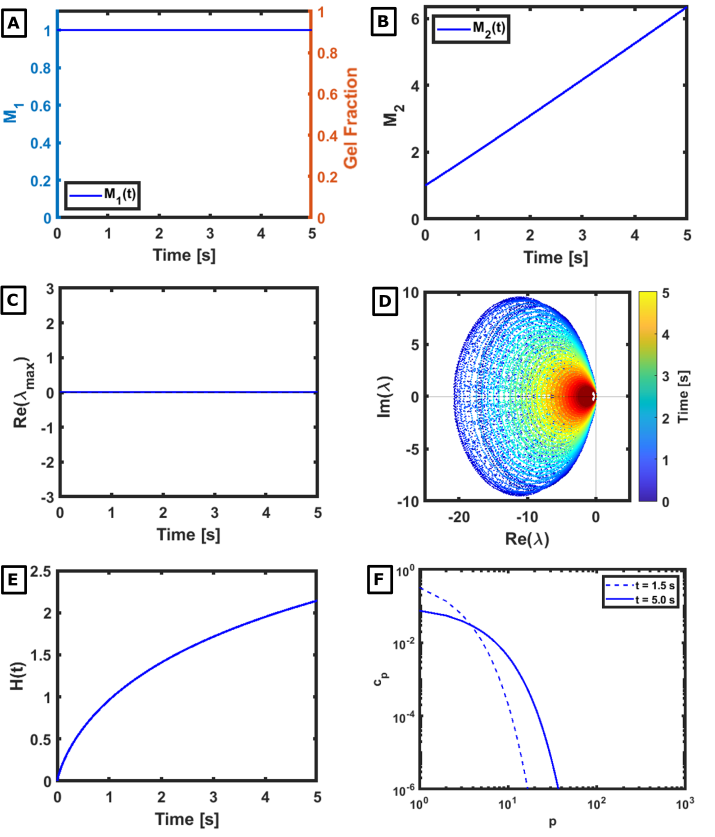}
    \caption{\textbf{Dynamics of the non-gelling Smoluchowski diffusion-limited kernel ($D_f=1.8$).} (A) Evolution of the first moment $M_1$ (blue), which remains strictly conserved, and the gel fraction (orange), which stays at zero. (B) The second moment $M_2$ grows monotonically but remains finite, contrasting with the divergence observed in gelling kernels. (C) The real part of the maximum eigenvalue, $\text{Re}(\lambda_{\text{max}})$, remains vanishingly small ($\approx 0$), indicating the persistence of dynamical stability. (D) The full instantaneous spectrum of the Jacobian in the complex plane; all eigenvalues are confined to the stable left half-plane ($\text{Re}(\lambda) \le 0$). (E) Evolution of the Shannon entropy $H(t)$. (F) Snapshots of the cluster size distribution $c_p$ for intermediate ($t=1.5$) and late ($t=5.0$) times.}
    \label{Fig_3}
\end{figure}

\section{Conclusion}

In summary, we demonstrate that kinetic gelation corresponds to a loss of local dynamical stability of the Smoluchowski flow, rather than being solely a moment-based singularity.\\
By bridging kinetic theory and non-linear dynamics, our approach provides a robust, model-independent diagnostic tool for predicting stability in complex aggregation systems.

\bibliographystyle{unsrt}
\bibliography{references}

\clearpage
\include{supporting_information}

\end{document}

%% file: Supporting_information.tex
\renewcommand{\thefigure}{S\arabic{figure}}
\setcounter{figure}{0}
\renewcommand{\thesection}{S\arabic{section}} 

\makeatletter
\begin{center}
    {\LARGE \bfseries SUPPORTING INFORMATION \par}
    \vskip 1em
    {\Large \bfseries IS GELATION A SINGULARITY OR A FLOW-INDUCED INSTABILITY? \par}
    \vskip 1em

    {\large
\textbf{Manuel Dedola}$^{\dag}$,
\textbf{Ludovico Cademartiri}$^{\dag *}$ \\[1em]
$^{\dag}$Department of Chemistry, Life Sciences and Environmental Sustainability,\\
University of Parma, Parco Area delle Scienze 17 A, Parma, Italy. \\[0.5em]
$^{*}$Corresponding author: \href{mailto:ludovico.cademartiri@unipr.it}{ludovico.cademartiri@unipr.it}
    \par}
    \vskip 1em
\end{center}
\makeatother

\RaggedRight

\setcounter{section}{0}
\setcounter{equation}{0}

\begin{figure}[H]
    \centering
    \includegraphics[width=0.9\linewidth]{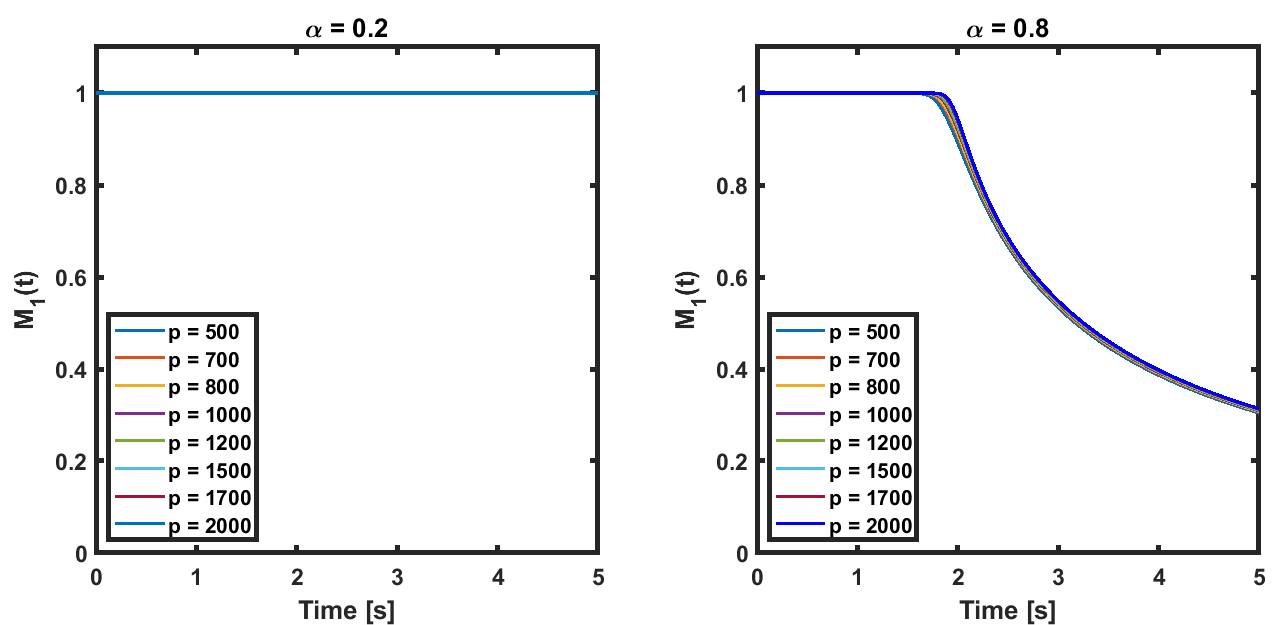}
    \caption{\textbf{First momentum $M_1(t)$ distribution for multiplicative kernel $K=(ij)^\alpha$}. $\alpha=0.2$ (A) and $\alpha=0.8$ (B).}
    \label{S1}
\end{figure}

\section{Linearization and Stability Analysis}

We investigate the local stability of the system by analyzing the evolution of infinitesimal perturbations $\delta\mathbf{c}$, which physically represent spontaneous thermodynamic fluctuations or finite-size deviations from the mean-field limit.\\
Let $\mathbf{c}(t)$ denote the state vector of the system, governed by the nonlinear differential equation $\dot{\mathbf{c}} = \mathbf{F}(\mathbf{c})$.

We introduce a small perturbation $\delta\mathbf{c}(t)$ such that the perturbed state is given by $\mathbf{c}(t) + \delta\mathbf{c}(t)$. Substituting this expression into the governing equation and performing a Taylor series expansion truncated at the first order, we obtain:

\begin{equation}
    \frac{d}{dt}(\mathbf{c} + \delta\mathbf{c}) \approx \mathbf{F}(\mathbf{c}) + \mathbf{J}(t) \, \delta\mathbf{c}
\end{equation}

By subtracting the equation for the unperturbed flow, $\dot{\mathbf{c}} = \mathbf{F}(\mathbf{c})$, from Eq. 1, we isolate the linear evolution equation for the perturbation:

\begin{equation}
    \dot{\delta\mathbf{c}} = \mathbf{J}(t) \, \delta\mathbf{c}
\end{equation}

Here, $\mathbf{J}(t)$ represents the $p \times p$ Jacobian matrix evaluated along the trajectory. Its elements, $J_{ij}(t)$, describe the sensitivity of the rate of change of the $i$-th component with respect to the $j$-th variable:

\begin{equation}
    J_{ij}(t) = \frac{\partial \dot{c}_i}{\partial c_j}
\end{equation}

Since the Jacobian $\mathbf{J}(t)$ is time-dependent, the stability analysis requires a local approach. We adopt the \textit{frozen-time analysis} framework, which assesses the instantaneous tendency of perturbations to grow or decay. At any fixed instant $t$, we treat the coefficients of the linearized system as momentarily constant (quasistationary approximation). Under this assumption, we seek instantaneous eigenmodes of the form:

\begin{equation}
    \delta\mathbf{c}(\tau) \sim e^{\lambda \tau} \mathbf{v}
\end{equation}

where $\tau$ represents a local time scale relative to the frozen instant $t$. Substituting this ansatz into the frozen-coefficient version of Eq. 2 leads to the instantaneous eigenvalue problem:

\begin{equation}
    \mathbf{J}(t) \mathbf{v} = \lambda(t) \mathbf{v}
\end{equation}

Here, $\lambda(t)$ represents the set of instantaneous eigenvalues and $\mathbf{v}(t)$ the corresponding eigenvectors. Since the general solution is a superposition of these modes, the long-term behavior of the perturbation is dominated by the eigenvalue with the largest real part. Consequently, if $Re(\lambda_{max}(t)) > 0$, the flow is locally expansive, providing a robust diagnostic for the onset of dynamical instability (gelation).